\documentclass[twocolumn,aps,prc,preprintnumbers,showpacs,showkeys,nofootinbib,superscriptaddress,fleqn,floatfix,tightenlines, 10pt]{revtex4-1}
\usepackage{amsmath,amsfonts,amssymb,amscd,amsxtra,amsthm}
\usepackage{natbib}
\usepackage{graphicx,subfigure}  % Include figure files
\usepackage{epstopdf}
\usepackage{dcolumn}  % Align table columns on decimal point
\usepackage{bm}          % bold math
\usepackage{slashed}
\usepackage[utf8]{inputenc}
\usepackage{CJK}
\usepackage{mathptmx}
\usepackage{mathrsfs}
\usepackage[normalem]{ulem} % \sout{old text} for strikeout
\usepackage[dvipsnames]{xcolor} % For blue in-text comments and
\usepackage{array}
\usepackage{multirow}

\begin{document}
%\preprint{Arxive}
\title{Pion parton distribution functions with the nonrelativistic constituent quark model}

\author{Qian Wu}
\affiliation{Institute of Modern Physics, Chinese Academy of Sciences, Lanzhou 730000, China}

\author{Chengdong Han}
\affiliation{Institute of Modern Physics, Chinese Academy of Sciences, Lanzhou 730000, China}
\affiliation{School of Nuclear Science and Technology, University of Chinese Academy of Sciences, Beijing 100049, China}

\author{Di Qing}
\affiliation{TRIUMF, Vancouver, BC V6T 2A3, Canada}

\author{Wei Kou}
\affiliation{Institute of Modern Physics, Chinese Academy of Sciences, Lanzhou 730000, China}
\affiliation{School of Nuclear Science and Technology, University of Chinese Academy of Sciences, Beijing 100049, China}

\author{Xurong Chen}~\email{xchen@impcas.ac.cn}
\affiliation{Institute of Modern Physics, Chinese Academy of Sciences, Lanzhou 730000, China}
\affiliation{School of Nuclear Science and Technology, University of Chinese Academy of Sciences, Beijing 100049, China}
\affiliation{Guangdong Provincial Key Laboratory of Nuclear Science, Institute of Quantum Matter,
South China Normal University, Guangzhou 510006, China}

\author{Fan Wang}~\email{fgwang@nju.edu.cn}
\affiliation{Department of Physics, Nanjing University, Nanjing 210093, China}

\author{Ju-Jun Xie}~\email{xiejujun@impcas.ac.cn}
\affiliation{Institute of Modern Physics, Chinese Academy of
	Sciences, Lanzhou 730000, China} \affiliation{School of Nuclear Science and Technology, University of Chinese Academy of Sciences, Beijing 100049, China} \affiliation{School of Physics and
	Microelectronics, Zhengzhou University, Zhengzhou, Henan 450001,
	China} \affiliation{Lanzhou Center for Theoretical Physics, Key
	Laboratory of Theoretical Physics of Gansu Province, Lanzhou
	University, Lanzhou 730000, China}

\begin{abstract}

We calculate the valence quark distribution functions of the $\pi$ meson using the non-relativistic chiral constituent quark model.
The $\pi$ wave function is obtained by solving the two-body
Schr\"odinger equation within the framework of constituent quark model.
We transform the $\pi$ wave function from the rest frame to the light cone or infinite momentum frame based on the Lorentz boost.
The valence quark distributions at the initial evolution scale are obtained. The QCD evolution are given with the DGLAP equations with parton-parton recombination corrections.
With tuning the valence up (down) quark mass to 70 MeV, the calculated valence up quark distributions at $Q^2=20$ GeV$^2$ are in good agreement with the E615 experimental data.
The structure functions $\rm{F}_2^\pi(x,Q^2)$ of pion are also calculated which consist with the H1 experimental data. The proposed mechanisms here could be also used to study other hadrons.

\end{abstract}
\maketitle
\section{Introduction}

The light pseudoscalar mesons have gained much interest in nuclear and hadron physics, mostly because they are Nambu-Goldstone bosons \cite{1Nambu1960,2Goldstone1962} which are good objects to understand the dynamical chiral symmetry breaking~\cite{3Roberts1998,4Roberts2020,5Roberts2020fewbody}. As the lightest meson, pion, the study of its internal structure helps us understand the emergence of hadron mass from quantum chromo dynamics (QCD). The cross section of deep inelastic lepton-hadron scattering experiments, exists as one of the major experiments to reveal the structure of the hadron,
can be interpreted as the quarks and gluons (partons) momentum fractions probability distribution. Therefore,
to get a better understanding of the non-perturbative dynamics in QCD, it is necessary to construct the connections of
gluon and quark parton distribution functions (PDFs) between experiment and theory \cite{6Holt2010}.

Experimental data on the pion PDFs are, however, scarce due to the lack of the fixed pion target. Currently the only valid data are inferred from Drell-Yan (DY) \cite{7DY} scattering on nuclear target
at CERN (NA3 and NA10) \cite{8DYstruc,9DYstruc2,10Betev1985}, Fermilab (E615) \cite{11Conway} and the leading-neutron deep elastic scattering of electron-proton collision at HERA (ZEUS and H1) \cite{12HERA2002,13HERA2010}. And the ZEUS and H1 measurements have better constrained
data of sea quarks and gluons distribution at small $x$. Recently, as an extensive study of the hadron structure
experiments at Jefferson Laboratory (JLAB), an experiment using the tagged deep-elastic scattering at the upgraded JLAB is approved \cite{14JLAB12-1,15JLAB12-2}, which accounts for the large Bjorken-$x$ behavior of pion.
Furthermore, new mesonic DY measurements at modern facilities may provide more valuable information on pion PDFs \cite{16newDY-1,17newDY-2}.

There are quite a lot of theoretical calculations of the pion PDFs, such as lattice QCD (LQCD) \cite{18LQCD-1,19LQCD-2,20LQCD-3}, Dyson-Schwinger Equations (DSE) \cite{21DSE-1,22DSE-2,23DSE-3}, light-front quantization \cite{24Lightfront}, maximum entrophy method (MEM) \cite{27han2020,27han2021} and the constituent quark model \cite{25CQM-PDF1,26CQM-PDF1994,27CQM-PDFmesoncloud}.
However, due to the nonperturbative nature in QCD, it remains one of the greatest issues in hadron physics
of building direct connections between the fundamental principles of QCD and the PDFs of the hadron.

The constituent quark model has been proved to well describe the low-energy hadronic phenomenology. In the low-energy region, the current quark is surrounded by a cloud of virtual quarks and gluons, which underlies the
large mass of the constituent quark. The constituent quark model is first applied into describing the heavy meson spectra of charmonium, which consists of a phenomenological confining interaction and a one gluon exchange
interaction \cite{28CQM1,29CQM2,30CQM3,31CQM4}. They successfully predict the charmed quark-antiquark states below the $D$-$D$ threshold and the decay widths of them.
Later, Isgur et al. successfully applied the constituent quark model into the baryon spectra \cite{32Isgur1,33Isgur2}.

As far as the light meson spectra, Vijande et al. reproduced almost all meson spectra with the SU(3) constituent quark model which proposed the GoldStone bosons exchange interactions~\cite{34CQM2005}, where they successfully reproduced the mass of pion with the constituent up (down) quark mass setting to 313 MeV which is about one third of the proton mass.
And in this work, we try to use this constituent quark model to describe the pion while the constituent up (down) quark mass is adjusted to 70 MeV which is half of the pion mass.
This will be discussed in the next section.

There are several calculations of the pion PDFs with the constituent quark model. In Ref. \cite{27CQM-PDFmesoncloud}, Watanabe et al. investigated the valence quark distributions of the pion in the framework of chiral
constituent quark model with considering the meson cloud effect. As we know, the light pseudoscalar meson are the Nambu-Goldstone bosons which couple to the hadron and develop the meson cloud.
With extracting the meson effect in pion, they concluded that the meson cloud effect cause a reduction of the valence
quark distribution and an enhancement at the small Bjorken $x$ regime. However, there is no comparison with the DY experimental data in their work.

Another important work in case of the constituent quark model and the hadron PDFs was done in Ref. \cite{26CQM-PDF1994}. With transforming the wave function in the rest frame into the light cone or infinity momentum frame (IMF), Dziembowski et al. calculated
the valence quark momentum-fraction probability distribution. Two different transformations were used in their work and one of them arises from the standard Lorentz transformation, which gives a narrow up quark distribution, located around $x$ $\sim0.3$.
Another transformation is referred from the so-called Licht-Pagamenta \cite{27LP-1975}
prescription and the so obtained valence quark distributions are in good agreement with the experimental data after the QCD evolution.
However, the Licht-Pagamenta prescription seems to violate its own assumption away from $x$ $\sim1/3$, which causes a number of technical and qualitative problems .

In this work, we aim to build connections between the constituent quark model and the pion PDFs.
First, we calculate the pion mass and its wave function within the SU(3) constituent quark model which including the confining interaction, one gluon exchange interaction and GoldStone bosons exchange interactions with the
Gaussian expansion method \cite{35hiyama1997ptp,35Hiyama2003GEM}.
And in order to connect the pion in the rest frame with the one in the infinity momentum frame (IMF), in which the deep inelastic scattering occurs, we do the inverse Lorentz transformation on the calculated wave function in the rest frame.
Using the wave function in the light cone, we calculate the valence quark distribution in the pion at the initial scale $Q_0^2$.
Then, with the QCD evolution done by the DGLAP equations \cite{36DGLAP-D,36DGLAP-GL,36DGLAP-AP}, we calculate the valence quark distributions at $Q^2=20$ GeV$^2$.
We also calculate the structure function ${\rm F}_2^{\pi}$ in the small x regime. Finally, we compare our calculated PDFs and structure functions with the experimental data.

This article is organized as follows. In Sec. \uppercase\expandafter{\romannumeral2}, we introduce the constituent quark model and the Gaussian expansion method, which we use to solve the Schr$\ddot{\rm o}$dinger equation.
We show how we transform our calculated wave function in the rest frame into the light cone in Sec. \uppercase\expandafter{\romannumeral3}. And in Sec. \uppercase\expandafter{\romannumeral4}, we briefly introduce the DGLAP equations
with parton-parton recombination corrections. The results and discussions are given in Sec. \uppercase\expandafter{\romannumeral5}. In the final section, we give a summary of this work.

\section{The constituent quark model}
We use the following chiral constituent quark model which consists of three parts: quark rest mass, kinetic energy and the potential energy \cite{34CQM2005}:
\begin{equation}
\begin{aligned}
H_{CQM} &=\sum_{i=1}^{2} m_{i}+\frac{p_{12}^{2}}{2 \mu_{12}} \\
&+V_{12}^{C}+V_{12}^{G}+\sum_{\chi=\pi, K, \eta} V_{12}^{\chi}+V_{12}^{\sigma},
\end{aligned}
\end{equation}
where the '$C$', '$G$' and '$\chi$' represent the quark confinement, one gluon exchange and the one Goldstone boson exchange, respectively.
The $m_i$ is the constituent quark mass, $\mu_{12}$ is the reduced mass of the two quarks and $p_{12}=(p_1-p_2)/2$. And the forms of these potentials are:
\begin{equation}
\begin{aligned}
V_{i j}^{C}=&=\left\{-a_{c}\left(1-e^{-\mu_{c} r_{i j}}\right)+\Delta\right\}\left(\overrightarrow{\lambda}_{i}^{c} \cdot \vec{\lambda}_{j}^{c}\right), \\
V_{i j}^{G}=& \frac{\alpha_{s}}{4} \lambda_{i}^{c} \cdot \lambda_{j}^{c}\left[\frac{1}{r_{i j}}-\frac{2 \pi}{3 m_{i} m_{j}} \sigma_{i} \cdot \sigma_{j} \delta\left(r_{i j}\right)\right], \\
\delta\left(r_{i j}\right)&= \frac{e^{-r_{i j} / r_{0}\left(\mu_{i j}\right)}}{4 \pi r_{i j} r_{0}^{2}\left(\mu_{i j}\right)}, \\
V_{i j}^{\pi}=& \frac{g_{c h}^{2}}{4 \pi} \frac{m_{\pi}^{2}}{12 m_{i} m_{j}} \frac{\Lambda_{\pi}^{2}}{\Lambda_{\pi}^{2}-m_{\pi}^{2}} m_{\pi} v_{i j}^{\pi} \sum_{f=1}^{3} \lambda_{i}^{f} \lambda_{j}^{f}, \\
V_{i j}^{K}=& \frac{g_{c h}^{2}}{4 \pi} \frac{m_{K}^{2}}{12 m_{i} m_{j}} \frac{\Lambda_{K}^{2}}{\Lambda_{K}^{2}-m_{K}^{2}} m_{K} v_{i j}^{K} \sum_{f=4}^{7} \lambda_{i}^{f} \lambda_{j}^{f}, \\
V_{i j}^{\eta}=& \frac{g_{c h}^{2}}{4 \pi} \frac{m_{\eta}^{2}}{12 m_{i} m_{j}} \frac{\Lambda_{\eta}^{2}}{\Lambda_{\eta}^{2}-m_{\eta}^{2}} m_{\eta} v_{i j}^{\eta} \\
& \times\left[\lambda_{i}^{8} \lambda_{j}^{8} \cos \theta_{P}-\lambda_{i}^{0} \lambda_{j}^{0} \sin \theta_{P}\right], \\
v_{i j}^{\chi}=& {\left[Y\left(m_{\chi} r_{i j}\right)-\frac{\Lambda_{\chi}^{3}}{m_{\chi}^{3}} Y\left(\Lambda_{\chi} r_{i j}\right)\right] \sigma_{i} \cdot \sigma_{j} }, \\
V_{i j}^{\sigma}=&-\frac{g_{c h}^{2}}{4 \pi} \frac{\Lambda_{\sigma}^{2}}{\Lambda_{\sigma}^{2}-m_{\sigma}^{2}} m_{\sigma} \\
& \times\left[Y\left(m_{\sigma} r_{i j}\right)-\frac{\Lambda_{\sigma}}{m_{\sigma}} Y\left(\Lambda_{\sigma} r_{i j}\right)\right], \\
\end{aligned}
\end{equation}
where $Y(x)=e^{-x}/x$ and $r_0(\mu_{ij})=s_o/\mu_{ij}$; $\sigma$ are the SU(2) Pauli matrices; $\lambda^f$ and $\lambda^c$ are the SU(3) flavor and color Gell-Mann matrices, respectively.
In addition, $g_{ch}^2/(4\pi)$ is the chiral coupling constant which is determined from the $\pi$-nucleon coupling and $\alpha_s$ is an effective
scale-dependent running coupling:
\begin{equation}
\alpha_{s}\left(\mu_{i j}\right)=\frac{\alpha_{0}}{\ln \left[\left(\mu_{i j}^{2}+\mu_{0}^{2}\right) / \Lambda_{0}^{2}\right]}.
\end{equation}

All parameters in the quark model are determined by fitting to the light meson spectra, which are listed in table.~\ref{tab:CQMPARA}. It should be noted that those parameters we used here are different with the ones used in
Ref.~\cite{34CQM2005}. The differences of the parameters originate from the mass of the constituent mass of the up (down) quark.
In this work, in order to obtain the appropriate valence quark distributions, which will be discussed later, we use the 70 MeV for the
constituent up (down) quark mass. It is nearly half of the $m_\pi$ while $m_u=313$ MeV is used in Ref.~\cite{34CQM2005} which is equal to one third of the proton mass.
Thus, the chiral coupling constant $g_{ch}$ should be modified with the different mass of the $m_u$. The coupling constant $g_{ch}$ is determined from the $\pi NN$ coupling constant through:
\begin{equation}
\frac{g_{c h}^{2}}{4 \pi}=\left(\frac{3}{5}\right)^{2} \frac{g_{\pi N N}^{2}}{4 \pi} \frac{m_{u, d}^{2}}{m_{N}^{2}},
\end{equation}
which is based on the assumption that the flavor SU(3) symmetry is broken only when the mass of the strange quark is changed.

\begin{table}[tbhp]
\caption{The constituent quark model parameters used in this work.}
\begin{tabular}{llr}
\hline Quark masses (MeV) & $m_{u}=m_{d}$ & 70 \\
                    & $m_s$    & 330 \\
\hline Goldstone bosons & $m_{\pi}$ & $0.70$ \\
$\left(\mathrm{fm}^{-1} \sim 200 \mathrm{MeV}\right)$ & $m_{\sigma}$ & $3.42$ \\
& $m_{\eta}$ & $2.77$ \\
& $m_{K}$ & $2.51$ \\
& $\Lambda_{\pi}=\Lambda_{\sigma}$ & $4.2$ \\
& $\Lambda_{\eta}=\Lambda_{K}$ & $5.2$ \\
\cline { 2 - 3 } & $g_{c h}^{2} /(4 \pi)$ & $0.027$ \\
& $\theta_{p}\left(^{\circ}\right)$ & $-15$ \\
\hline Confinement & $a_{c}\left(\mathrm{MeV}\right)$ & 590 \\
& $\mu_{c}\left(\mathrm{fm}^{-1}\right)$ & 0.70 \\
& $\Delta(\mathrm{MeV})$ & $305.41$ \\
\hline OGE & $\alpha_{0}$ & $1.532$ \\
& $\Lambda_{0}\left(\mathrm{fm}^{-1}\right)$ & $0.033$ \\
& $\mu_{0}(\mathrm{MeV})$ & $36.98$ \\
& $s_{0}(\mathrm{MeV})$ & $39.41$ \\
\hline
\end{tabular}
\label{tab:CQMPARA}
\end{table}

With the parameters shown in table.~\ref{tab:CQMPARA}, one can obtain the light meson spectra as in Table. \ref{tab:level}, compared with the experimental data~\cite{PDG}.
Note that the mass of the $\eta$ meson can't be reproduced with our present model which may need to consider the influence of the gluons or sea quarks.
Moreover, in order to reproduce the mass of the $K$ meson, the mass of the constituent strange quark is fixed to 330 MeV.
\begin{table}[tbh]
\caption{Mass spectra of light mesons. The units are in MeV.}
\begin{tabular}{p{2cm}<{\centering}p{2.5cm}<{\centering}p{2cm}<{\centering}}
  \hline
 Meson   & Theory  & Experiment  \\
 \hline
 $\pi$    & 139              & 139.57   \\
   $K$    & 488            & 497.61   \\
 $\rho$   & 776              & 775.26   \\
 $\omega$ & 761              & 782.65   \\
 \hline
\end{tabular}
\label{tab:level}
\end{table}

In order to obtain the mass spectra and the wave function of $\pi^+$, we solve the following Schr\"odinger equation:
\begin{equation}
H_{CQM} =E^{I J} \Psi_{M_{I} M_{J}}^{I J}(1,2).
\end{equation}
Here, $H_{CQM}$ is the chiral quark model hamiltonian and 1, 2 are the particle labels which represent up quark and anti-down quark in the $\pi^+$, respectively. The quantum number $I(M_I)$ and $J(M_J)$ represent the isospin and angular momentum
and their z-components, respectively. To solve the Schr\"odinger equation, we apply the Gaussian expansion method \cite{35Hiyama2003GEM}. Then, we write the wave function as:
\begin{equation}
\begin{aligned}
&\Psi_{M_{I} M_{J}}^{I J}(1,2) \\
&=\sum_{n,l,s,I} C_{n,l,s,I} \left[\psi_{n l}^{G}(\mathbf{r}) \chi_{s}(1,2)\right]_{M_{J}}^{J} \omega^{c}(1,2) \phi_{M_{I}}^{I}(1,2),
\end{aligned}
\end{equation}
where the spatial part is expanded with the Gaussian forms:
\begin{equation}
\psi_{n l m}^{G}(\mathrm{r}) =N_{n l} r^{l} e^{-\nu_{n} r^{2}} Y_{l m}(\hat{\mathrm{r}}),
\end{equation}
with the Gaussian parameters choosing in the Geometric progression:
\begin{equation}
\nu_{n}=\frac{1}{r_{n}^{2}}, \quad r_{n}=r_{1} a^{n-1}, \quad a=\left(\frac{r_{n_{\max }}}{r_{1}}\right)^{\frac{1}{n_{\max -1}}}.
\end{equation}
The eigen energies and the coefficients $C_{n,l,s,I}$ are obtained with applying the Rayleigh-Ritz variational method.
%------------------------Lorentz------------------------------------
\section{The wave function in the light cone}
As we know, in the deep inelastic scattering, the hadron can be regarded as moving with an infinite momentum. And L.Susskind proposed that the IMF limiting
procedure is essentially a change from the laboratory time and z coordinates to the light cone time and space coordinates \cite{Susskind}. Here we use Susskind's method to transform our
calculated wave function from the rest frame into the light cone.

In the IMF, the hadron is moving with an infinite four-momentum $(E,P)$ in the z-direction. And for a quark in the pion in the rest frame, the four-momentum are $(k_0,\emph{k})$. Thus, the quark four-momentum
$(p_0,\emph{p})$ is given by the inverse Lorentz boost $p=L(p \leftarrow k)k$, i.e.,
\begin{equation}
\begin{aligned}
p_{0} &=\frac{E}{M}k_{0}-\frac{P}{M} k_{z}, \\
p_{z} &=\frac{E}{M} k_{z}-\frac{P}{M} k_{0}, \\
p_{\perp} &=k_{\perp} .
\end{aligned}
\end{equation}

According to Ref.~\cite{26CQM-PDF1994}, the longitudinal momentum of the quark in the IMF, $k_z$, can be expressed in terms of a fraction $\zeta$,
\begin{equation}
k_{z}=\zeta P \text { with } \sum \zeta=1 .
\end{equation}
Then, with $P\rightarrow \infty$ and $\zeta$ fixed positive, the quark on-shell energy is expanded as:
\begin{equation}
\begin{aligned}
k_{0} &=\sqrt{\zeta^{2} P^{2}+k_{\perp}^{2}+m^{2}} \\
&=\zeta P+\frac{k_{\perp}^{2}+m^{2}}{2 \eta P}+O\left(P^{-3}\right),
\end{aligned}
\end{equation}
and similarly, the pion energy is expanded as,
\begin{equation}
E=\sqrt{P^{2}+M^{2}}=P+\frac{M^{2}}{2 P}+O\left(P^{-3}\right) .
\end{equation}
Then, the inverse Lorentz boost of Eq. (9) is obtained as:
\begin{equation}
\begin{aligned}
p_{0} &=\frac{1}{2}\left(\zeta M+\frac{p_{\perp}^{2}+m^{2}}{\eta M}\right), \\
p_{z} &=\frac{1}{2}\left(\zeta M-\frac{p_{\perp}^{2}+m^{2}}{\eta M}\right), \\
p_{\perp} &=k_{\perp} .
\end{aligned}
\end{equation}
In fact, the Lorentz boost in Eqs. (9) and (13) gives relations between the rest frame $p$ and the IMF variables $(\eta,{k}_{\perp})$. And noticing in the IMF limit,
the longitudinal fraction $\zeta=k_z/P$ can be replaced with the light cone fraction $x=k^+/P^+$, i.e.,
\begin{equation}
\begin{aligned}
p_{0} &=\frac{1}{2}\left(p^{+}+p^{-}\right)=\frac{1}{2}\left(x M+\frac{p_{\perp}^{2}+m^{2}}{x M}\right), \\
p_{z} &=\frac{1}{2}\left(p^{+}-p^{-}\right)=\frac{1}{2}\left(x M-\frac{p_{\perp}^{2}+m^{2}}{x M}\right), \\
p_{\perp} &=k_{\perp} .
\end{aligned}
\end{equation}

With applying Eq. (14) into the calculated momentum wave function of pion, $\varphi^{\pi^+}(\bf{p_1},\bf{p_2})$, we obtain the light cone momentum wave function as follows:
\begin{equation}\label{lightcone}
\varphi^{\pi^+}_{LC}(x_1,x_2,\vec{p_{\perp1}},\vec{p_{\perp2}})=N^{\pi^+}\varphi^{\pi^+}(\bf{p_1},\bf{p_2})/\sqrt{x_1x_2}.
\end{equation}
Here the $1/\sqrt{x_1x_2}$ is the light cone factor and $N^{\pi^+}$ is the normalization factor. It should be noted that the normalization condition is broken when Eq. (14) is applied into the wave function $\varphi^{\pi^+}(\bf{p_1},\bf{p_2})$.
Thus the light cone wave function need to be re-normalized again and the normalization factor $N^{\pi^+}$ is then determined by:
\begin{equation}
\int[d x]\left[d^{2} \vec{p}_{\perp}\right]\left|\varphi_{L C}\right|^{2}=1,
\end{equation}
with
\begin{equation}
\begin{aligned}
{[d x]=\prod d x_i \delta\left(\sum x_i-1\right) }, \\
{\left[d^{2} \vec{p}_{\perp}\right]=\prod\left[d^{2} \vec{p_i}_{\perp}\right] \delta\left(\sum \vec{p_i}_{\perp}\right) }.
\end{aligned}
\end{equation}

With these above ingredients, the wave function in the rest frame calculated with the constituent quark model can be easily transformed into a valence component of the light cone wave function.

\section{DGLAP equations with parton-parton recombination corrections}
In order to connect the nonperturbative result with the experimental measurements in the asymptotic region, the QCD evolution equations are used. Among them,
the Dokshitzer-GribovLipatov-Altarelli-Parisi (DGLAP) equations are a widely used tool to describe $Q^{2}$ dependence of quark and gluon distributions \cite{36DGLAP-AP,36DGLAP-D,36DGLAP-GL,36DGLAP-GL}.
They are derived from perturbative QCD within the quark-parton model and provide the picture of parton evolution with $Q^2$. The DGLAP equations have the form as:
\begin{equation}
\begin{aligned}
&\frac{\mathrm{d}}{\mathrm{d} \ln Q^{2}}\left(\begin{array}{c}
f_{\mathrm{q}_{i}}\left(x, Q^{2}\right) \\
f_{\mathrm{g}}\left(x, Q^{2}\right)
\end{array}\right)\\
&=\Sigma_{j} \int_{x}^{1} \frac{\mathrm{d} z}{z}\left(\begin{array}{cc}
P_{\mathrm{q}_{i} \mathrm{q}_{j}}(z) & P_{\mathrm{q}_{i} \mathrm{~g}}(z) \\
P_{\mathrm{gq}_{j}}(z) & P_{\mathrm{gg}}(z)
\end{array}\right) \times\left(\begin{array}{c}
f_{\mathrm{q}_{j}}\left(x, Q^{2}\right) \\
f_{\mathrm{g}}\left(x, Q^{2}\right)
\end{array}\right)
\end{aligned}
\end{equation}
where $P_{\mathrm{q}_{i}\mathrm{q}_{j}}$, $P_{\mathrm{q}_{i} \mathrm{~g}}$, $P_{\mathrm{gq}_{j}}$ and $P_{\mathrm{gg}}$ are the parton splitting functions \cite{36DGLAP-GL}.

One important correction to the DGLAP equations is the so-called parton-parton recombination effect.
The parton-parton recombination effect was first put forward by Gribov, Levin and Ryskin (GLR) \cite{37DGLAP-GLR}, then followed by Mueller, Qiu evaluating the recombination coefficients with the covariant field theory \cite{38DGLAP-MQ}.
The gluon number density grows rapidly at small $x$ and
the quanta of partons overlap spatially. Therefore, the effect of the parton-parton interaction is no longer ignorable and it is expected to stop the increase of the cross sections near the unitarity limits.
The parton-parton recombination corrections including gluon-gluon, quark-gluon, quark-quark were derived by Zhu, Ruan, and Shen with the time-ordered perturbative theory \cite{39DGLAP-ZRS1,39DGLAP-ZRS2,39DGLAP-ZRS3,40wang2017}.
Then, the DGLAP equations with parton-parton recombination corrections used in this work are:
\begin{equation}
Q^{2} \frac{d x f_{q_{i}}\left(x, Q^{2}\right)}{d Q^{2}}=\frac{\alpha_{s}\left(Q^{2}\right)}{2 \pi} P_{q q} \otimes f_{q_{i}},
\end{equation}
for valence quark distributions,
\begin{equation}
\begin{aligned}
Q^{2}& \frac{d x f_{\bar{q}_{i}}\left(x, Q^{2}\right)}{d Q^{2}}= \frac{\alpha_{s}\left(Q^{2}\right)}{2 \pi}\left[P_{q q} \otimes f_{\bar{q}_{i}}+P_{q g} \otimes f_{g}\right] \\
&-\frac{\alpha_{s}^{2}\left(Q^{2}\right)}{4 \pi R^{2} Q^{2}} \int_{x}^{1 / 2} \frac{d y}{y} x P_{g g \rightarrow \bar{q}}(x, y)\left[y f_{g}\left(y, Q^{2}\right)\right]^{2} \\
&+\frac{\alpha_{s}^{2}\left(Q^{2}\right)}{4 \pi R^{2} Q^{2}} \int_{x / 2}^{x} \frac{d y}{y} x P_{g g \rightarrow \bar{q}}(x, y)\left[y f_{g}\left(y, Q^{2}\right)\right]^{2},
\end{aligned}
\end{equation}
for sea quarks distributions and
\begin{equation}
\begin{aligned}
Q^{2} &\frac{d x f_{g}\left(x, Q^{2}\right)}{d Q^{2}}= \frac{\alpha_{s}\left(Q^{2}\right)}{2 \pi}\left[P_{g q} \otimes \Sigma+P_{g g} \otimes f_{g}\right] \\
&-\frac{\alpha_{s}^{2}\left(Q^{2}\right)}{4 \pi R^{2} Q^{2}} \int_{x}^{1 / 2} \frac{d y}{y} x P_{g g \rightarrow g}(x, y)\left[y f_{g}\left(y, Q^{2}\right)\right]^{2} \\
&+\frac{\alpha_{s}^{2}\left(Q^{2}\right)}{4 \pi R^{2} Q^{2}} \int_{x / 2}^{x} \frac{d y}{y} x P_{g g \rightarrow g}(x, y)\left[y f_{g}\left(y, Q^{2}\right)\right]^{2},
\end{aligned}
\end{equation}
for gluon distribution.
Here $P_{q q}$, $P_{q g}$, $P_{g q}$, $P_{g g}$ are the standard parton splitting kernels, and $P_{g g \rightarrow \bar{q}}$, $P_{g g \rightarrow g}$ are the gluon-gluon recombination coefficients.
The factor $1 /\left(4 \pi R^{2}\right)$ is from the normalization of the two-parton densities and $R$ is the correlation length of two interacting partons. The value of $R$ is related to the radii of the hadrons which is smaller than the hadron radius in most cases \cite{38DGLAP-MQ}. In this analysis for $\pi^+$, we take $R= 2.39$ ${\rm GeV}^{-1}$ which is determined with the global fit to the pion experimental data~\cite{27han2021}.

In addition, the $\Sigma$ in Eq. (21) is defined as
\begin{eqnarray}
\Sigma\left(x, Q^{2}\right) \equiv \sum_{j} f_{q_{j}}\left(x, Q^{2}\right)+\sum_{i}\left[f_{q_{i}}\left(x, Q^{2}\right)+f_{\overline{q}_{i}}\left(x, Q^{2}\right)\right].
\end{eqnarray}

\section{Results and Discussions}

\begin{figure}[htbp]
\setlength{\abovecaptionskip}{0.cm}
\setlength{\belowcaptionskip}{-0.cm}
\centering
\includegraphics[width=0.5\textwidth]{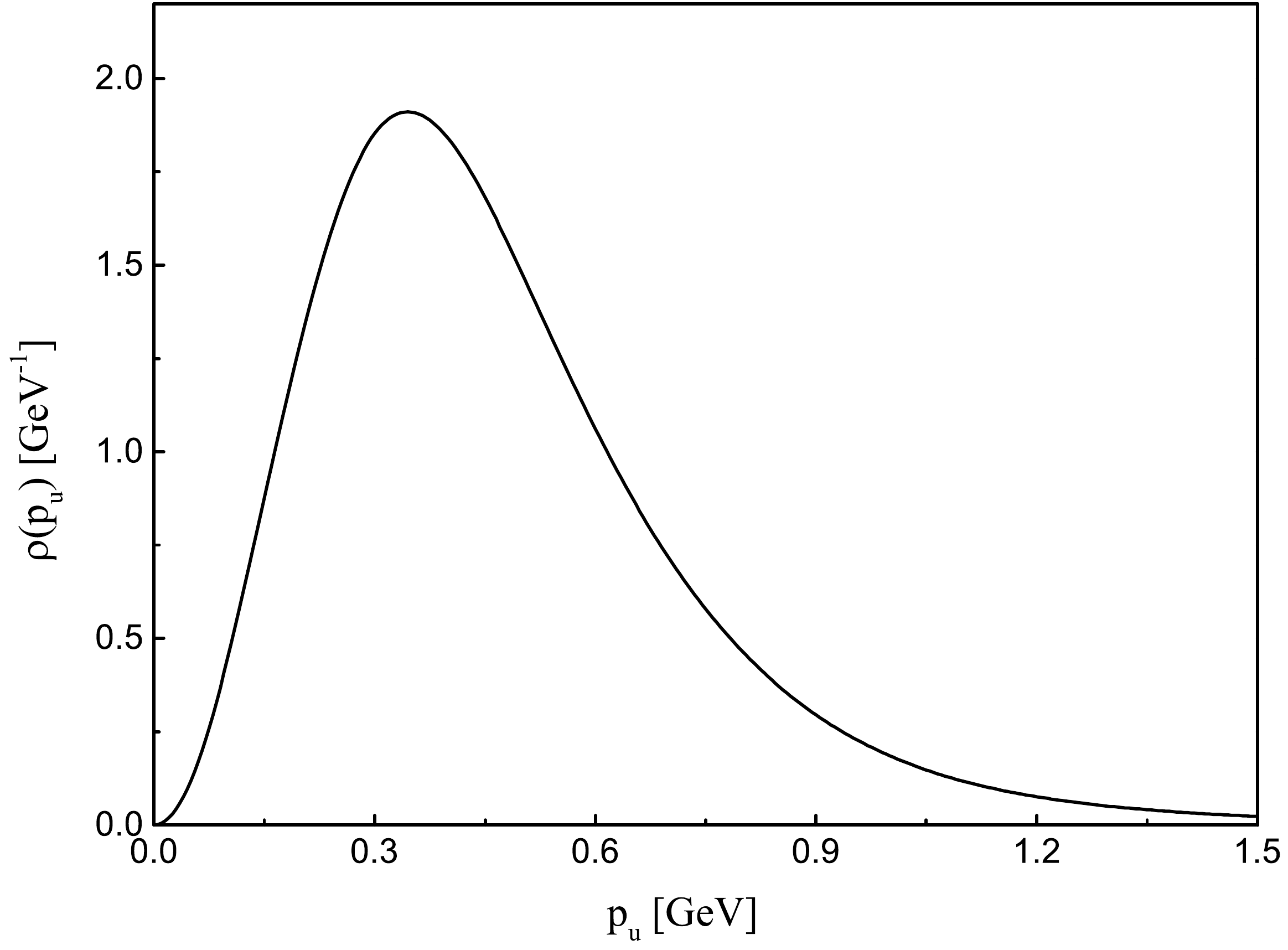}
\caption{
The momentum distribution $\rho(p_u)$ as a function of the relative momentum between the up quark and the anti-down quark.
}
\label{fig:momentum}
\end{figure}

%------------------------wave function------------------------------------------------
First, we obtain the conventional spatial wave function $\Psi_{M_{I} M_{J}}^{I J}(1,2)$ in the rest frame, then we transform it into the momentum space, we get the wave function in the momentum space, $\varphi^{\pi^+}(\bf{p_1},\bf{p_2})$.
We define the quark momentum distribution as follows:
\begin{equation}\label{densitymomentum}
\rho(p_u)=\langle\varphi^{\pi^+}(\bf{p_1},\bf{p_2})|\delta(p-p_u)|\varphi^{\pi^+}(\bf{p_1},\bf{p_2})\rangle,
\end{equation}
where $p_u$ is the relative momentum between the up quark and the anti-down-quark and $\varphi(\pi^+)$ is the momentum wave function of the $\pi^+$.
In Fig. \ref{fig:momentum}, we show the momentum distribution of up quark in the $\pi^+$.
%\begin{equation}\label{density}
%\rho(r_{1-c})=\int \psi(\emph{r_{1-c}})d\hat{r_{1-c}}
%\end{equation}

Then, the up quark momentum distribution of the pion in the light cone is given as:
\begin{equation}\label{densitymomentumLC}
\rho(x_u,\vec{p}_{\perp,u})=\langle\varphi^{\pi^+}_{LC}|\delta(x-x_u)\delta^2(p-\vec{p}_{\perp,u})|\varphi^{\pi^+}_{LC}\rangle,
\end{equation}
where $\varphi^{\pi^+}_{LC}$ is the calculated light cone wave function of the $\pi^+$, given in Eq. (15).
As mentioned in Eqs. (10) and (11), we treat the $p^2_\perp/P$ as an infinitesimal value. In order to check its validity, we calculate the average up quark transverse momentum $\langle p^2_\perp\rangle$ in the pion,
\begin{equation}
\left\langle p_{\perp}^{2}(x)\right\rangle=\frac{\int p_{\perp}^{2} d^{2} p_{\perp} \rho\left(x, \mathbf{p}_{\perp}\right)}{u_{v}(x)}.
\end{equation}

\begin{figure}[htbp]
\setlength{\abovecaptionskip}{0.cm}
\setlength{\belowcaptionskip}{-0.cm}
\centering
\includegraphics[width=0.5\textwidth]{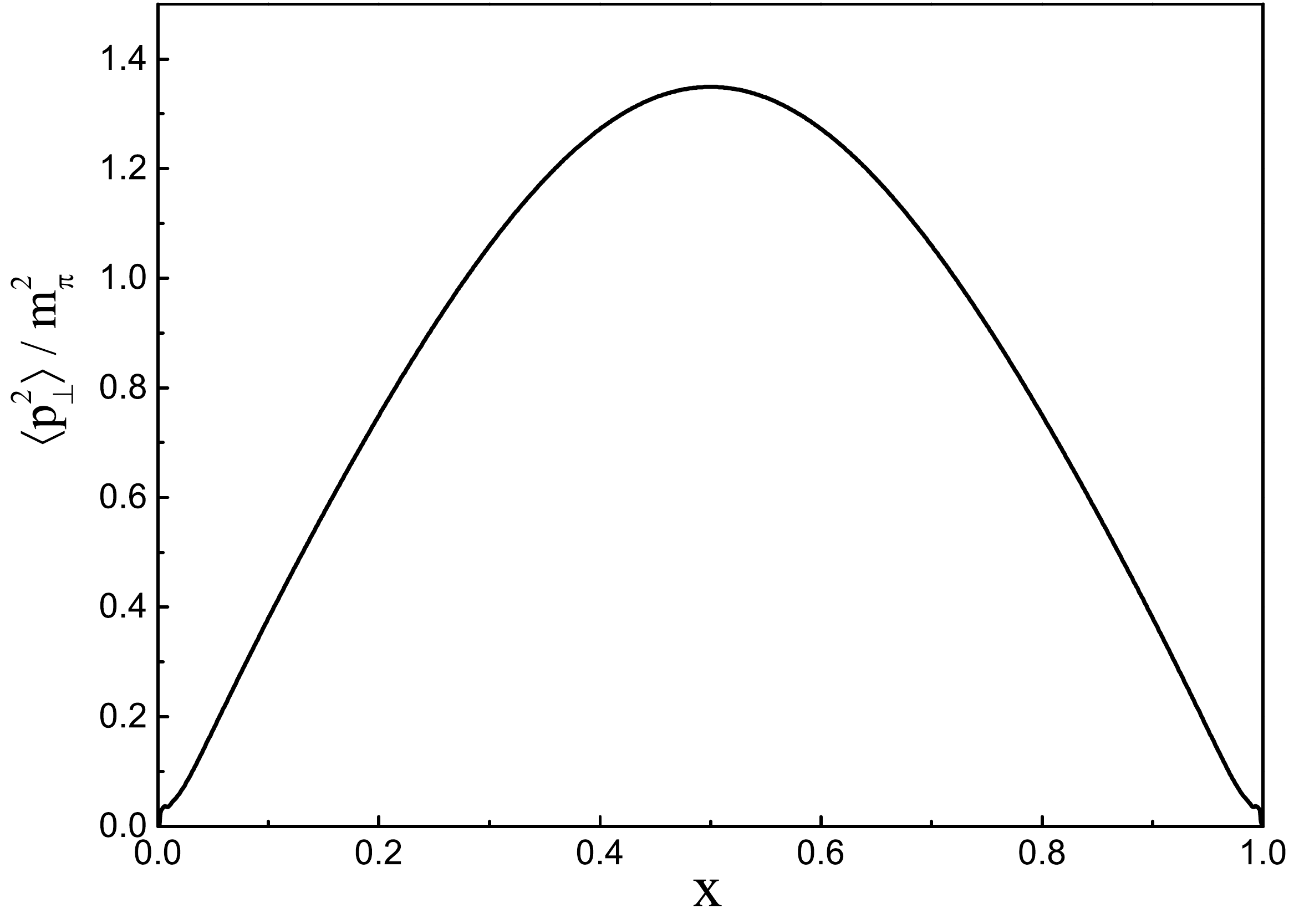}
\caption{The up quark transverse momentum distribution $\langle\rm{p}^2_\perp\rangle$ in the $\pi^+$.}
\label{fig:pion-tmd-1}
\end{figure}

In Fig.~\ref{fig:pion-tmd-1}, we show the obtained average transverse momentum $\langle p^2_\perp\rangle$ as a function of $x$. One can see that the transverse momentum in the pion are at most $1.3 m^2_\pi$, which is extremely small compared with the infinite momentum $P$ in the real experimental situation. Therefore, we conclude that our approximations in Eqs.~(10) and (11) are reliable. Besides, the transverse momentum vanishes around $x=0$ and $x=1$ since it's on the kinematical regions.

In order to obtain valence quark distribution with respect to the Bjorken variable $x$, we should do the integration over the transverse momentum $\vec{p_{\perp,u}}$:
\begin{equation}
u_{v}(x)=\int d^{2} \vec{p}_{\perp,u} \rho(x_u,\vec{p}_{\perp,u}).
\end{equation}

Since the $\pi^+$ meson consists of one up quark and one anti-down-quark, the valence quark distribution $u_v(x)$ above must satisfy the valence quark number sum rule and the momentum sum rule, i.e.,
\begin{equation}
\begin{array}{r}
\int_{0}^{1} u_{\mathrm{V}}^{\pi}\left(x, Q_{0}^{2}\right) d x=1, \\
\int_{0}^{1} \bar{d}_{\mathrm{V}}^{\pi}\left(x, Q_{0}^{2}\right) d x=1, \\
\int_{0}^{1} x\left[u_{\mathrm{V}}^{\pi}\left(x, Q_{0}^{2}\right)+\bar{d}_{\mathrm{V}}^{\pi}\left(x, Q_{0}^{2}\right)\right] d x=1.
\end{array}
\end{equation}

\begin{figure}[htbp]
\setlength{\abovecaptionskip}{0.cm}
\setlength{\belowcaptionskip}{-0.cm}
\centering
\includegraphics[width=0.45\textwidth]{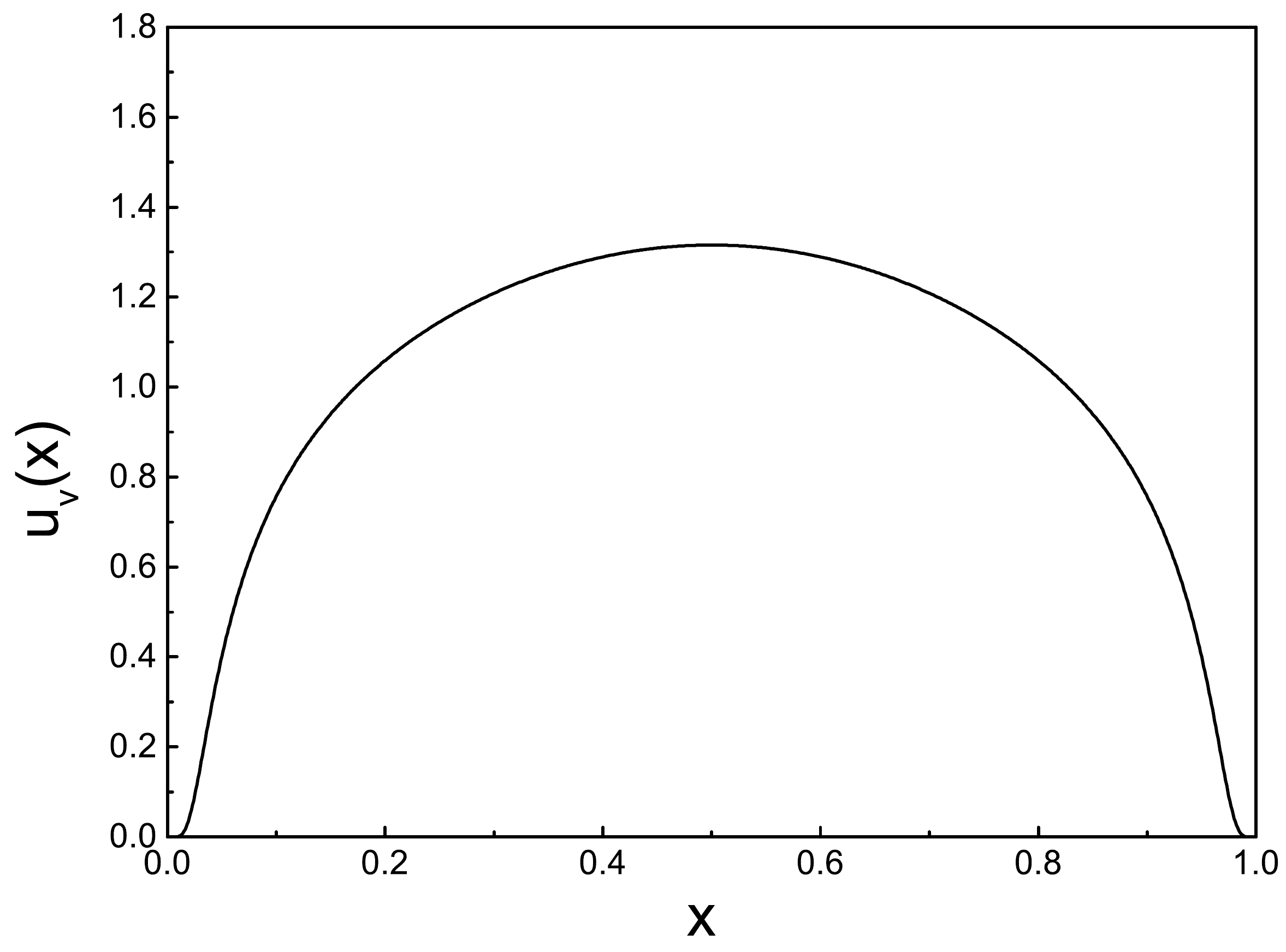}
\caption{
The valence up quark distribution of the $\pi^+$ based on the light cone wave function.
}
\label{fig:pion-1}
\end{figure}

In Fig.~\ref{fig:pion-1}, we depict the valence quark distribution of up quark in the $\pi^+$.~\footnote{The valence up quark and anti-down-quark distributions are same within our model.}
One may notice that, in the small $x<0.1$ and large $x>0.9$ regions, the valence quark distributions have an asymptotic behavior.
It comes from the damping behavior of the Gaussian type wave function, $e^{-r^2}$, which we use in Eq. (7). Accordingly, we use the following form to parameterize the valence quark distributions:
\begin{equation}\label{parapdf}
u_v(x)=Ax^B(1-x)^B\times g(x),
\end{equation}
where $g(x)$ reflects the asymptotic behavior of our calculated valence quark distributions, which is set as $g(x)=e^{-C(\frac{1}{x^2}+\frac{1}{(1-x)^2})}$ accounts for the asymptotic behavior when $x$ is close to 0 or 1. We take $A=2.58365$, $B=0.47900$ and $C=0.00087$.

%--------------------pion---Q_0---PDF------------------------------------------------

\begin{figure}[htbp]
\setlength{\abovecaptionskip}{0.cm}
\setlength{\belowcaptionskip}{-0.cm}
\centering
\includegraphics[width=0.5\textwidth]{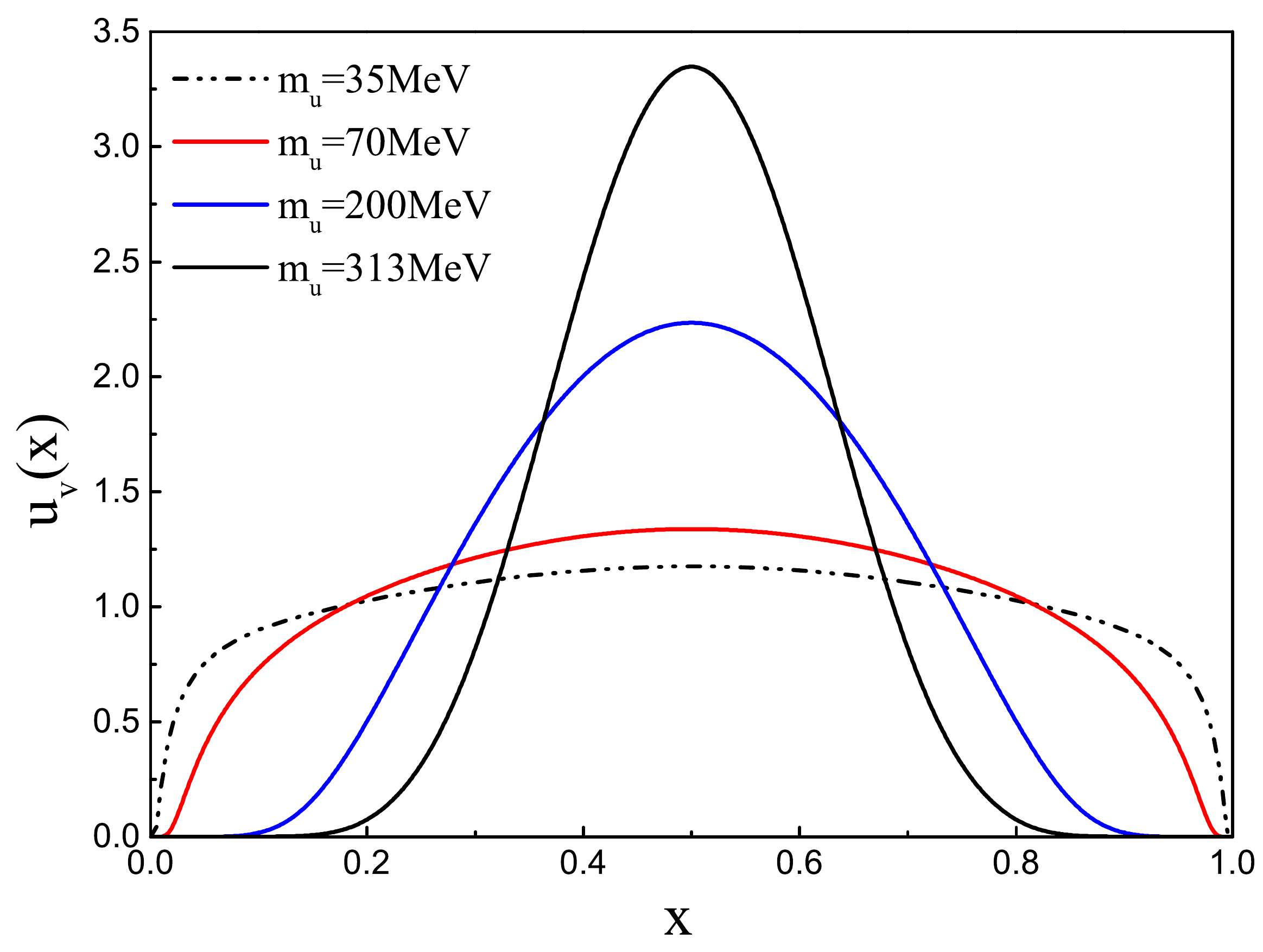}
\caption{
The valence quark distribution of the $\pi^+$ based on the light cone wave function with different mass of the light quark.
}
\label{fig:pion-m-1}
\end{figure}

Next, we consider the up quark mass effects on its distribution. The wave functions of the pion with different mass of the up quark are calculated. Then, we obtain the relation between the valence quark distributions of the pion and the mass of the up quark.
In Fig.~\ref{fig:pion-m-1}, we give the pion PDFs with several different masses of the up quark. It is found that the up quark distribution is broader with a smaller up quark mass.
If we choose a large constituent quark mass such as 313 MeV, we will obtain a narrow
up quark distribution which certainly contradicts with the experimental data after the QCD evolution. The reason for this relation between the quark mass and the valence quark distributions needs to further study.

%--------------------pion---Q_0---PDF---mass dependence---------------------------------------------

In Ref.~\cite{DSEMP}, within the Dyson-Schwinger equations, it gives a relation between the dressed quark mass in the propagator and its momentum, $M(p)$.
It shows that the dressed quark mass is turning small while the momentum of the quark is growing large. They think that the quark mass is not a fixed value while it
depends on the momentum scale. Here, the constituent quark, in some sense, appears like the dressed quark since it also behaves like the current quark surrounding by a could of virtual quarks and gluons. Besides, as mentioned in Sec. \uppercase\expandafter{\romannumeral2}, in order to obtain an appropriate valence quark distribution at the initial scale, we set the constituent quark mass equal to the half mass of the pion, 70 MeV, instead of using the one third of the proton mass, 313 MeV.

\begin{figure}[htbp]
\setlength{\abovecaptionskip}{0.cm}
\setlength{\belowcaptionskip}{-0.cm}
\centering
\includegraphics[width=0.5\textwidth]{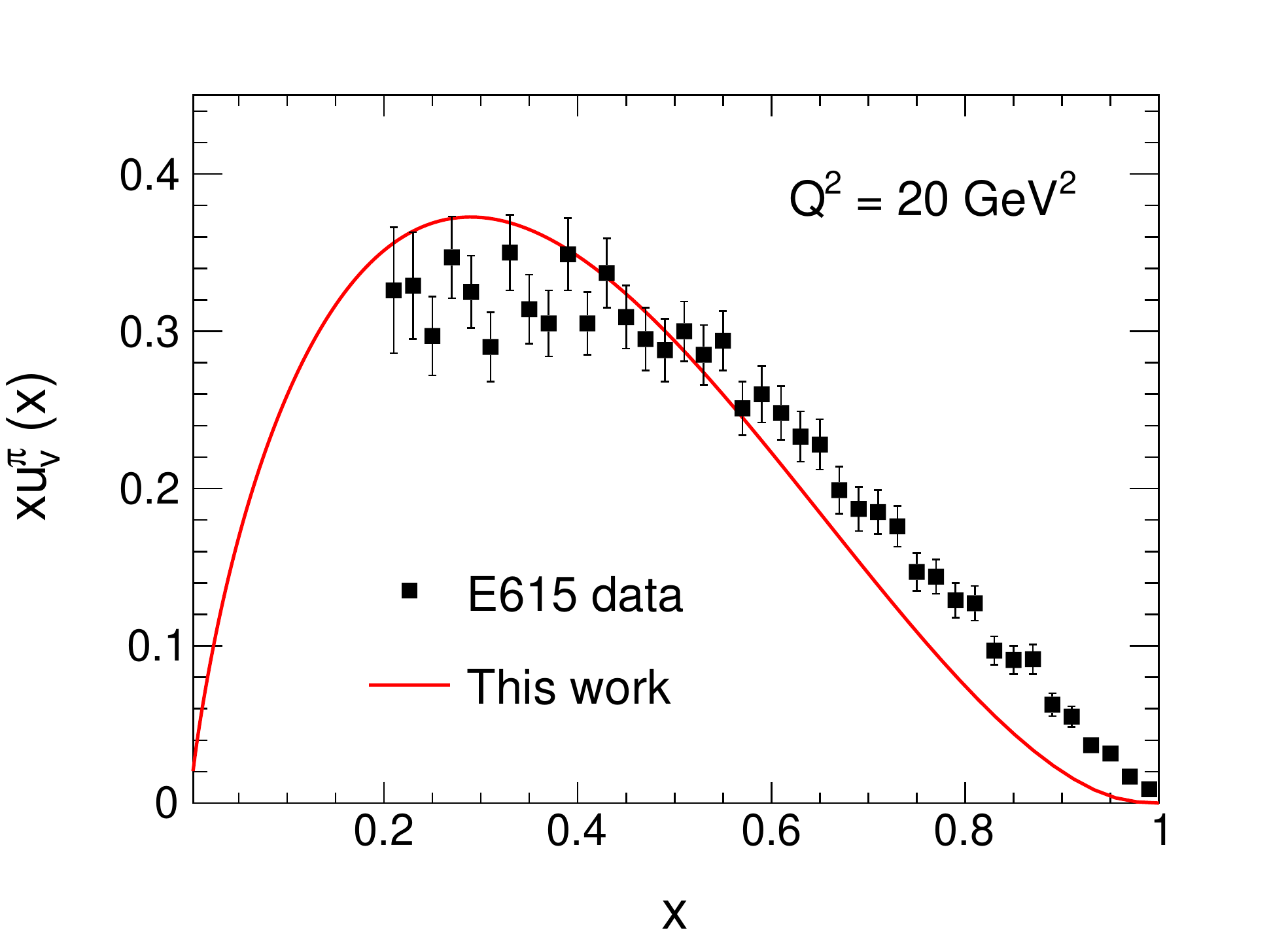}
\caption{
Comparisons between our calculated valence quark distribution at $Q^2=20$ GeV$^2$ and the E615 experimental data \cite{11Conway}.
}
\label{fig:pion-013}
\end{figure}

On the other hand, the initial evolution scale is taken as $Q_0^2=0.13$ ${\rm GeV}^2$~\cite{27han2021}, which is given by the global QCD analysis of the pion experimental data. In Fig. \ref{fig:pion-013}, we depict the valence quark distributions at $Q^2=20$ ${\rm GeV}^2$ by calculating the
DGLAP QCD evolution equations with nonlinear corrections. For comparison, the Drell-Yan data from Fermilab E615 experiments is also depicted \cite{11Conway}. We find that our calculated pion valence quark distributions are in good agreement with
the experimental data.

%------------------------Q=20GeV EVOLVED PDF-----------------------------------------------

\begin{figure*}[htbp]
\setlength{\abovecaptionskip}{0.cm}
\setlength{\belowcaptionskip}{-0.cm}
\begin{center}
\centerline{\includegraphics[width=16.0 cm,height=12.0 cm]{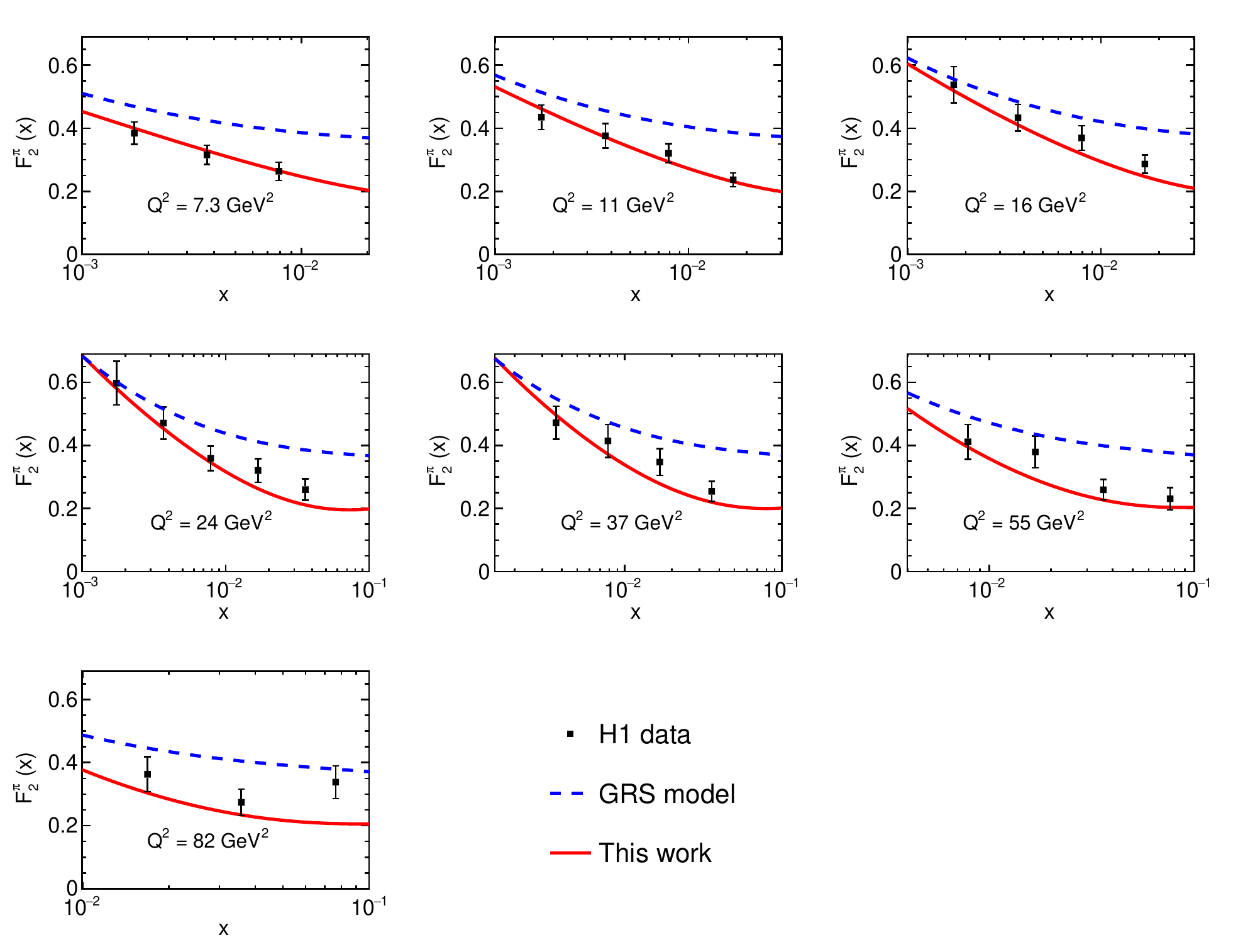}}
\end{center}
\caption{Comparisons between our predicted structure function $F_2^{\pi}(x,Q^2)$ and the H1 data \cite{13HERA2010} and GRS model predictions \cite{GRSmodel}.}
\label{fig:pion-013-2}
\end{figure*}

Finally, we turn to the $\pi^+$ structure function. Despite there is no PDFs in the small $x$ region for the pion,
with measuring the leading neutron tagged deep inelastic electron-proton scattering and using one-pion exchange model, H1 Collaboration at HERA extracted the pion structure functions at small $x$.
In Fig.~\ref{fig:pion-013-2}, we show the comparisons between our calculated pionic structure functions and the H1 experimental data \cite{13HERA2010}, and the GRS model predictions \cite{GRSmodel}.
One finds that in the small $x$ region, our pionic structure function results are in good agreement with the experimental data.
Hence, combing with all the experimental results in the small $x$ region and large $x$ region, our calculated valence quark distributions consist with the experimental data.

%------------------------Q=20GeV EVOLVED PDF---mass dependence--------------------------------------------

\section{Summary}

We calculate the pion mass and its wave function in the rest frame with the SU(3) constituent quark model. The mass of the constituent light quarks is set to 70 MeV. With appropriate adjustments of the quark model parameters, we reproduce the pion mass and its charge radius. We transform the calculated wave function from the rest frame into the light cone with the Lorentz boost. Then, we obtain the parton distribution of the valence up quark in the pion at initial scale.

With the global analysis of the current available data, our initial hadronic scale $Q_0^2$ is 0.13 GeV$^2$. Using the DGLAP equations with parton-parton recombination corrections, we obtain the valence quark distributions at $Q^2=20$ ${\rm GeV}^2$. Our results are in agreement with the E615 experimental data~\cite{11Conway}. Furthermore, we calculated also the $F_2^{\pi}$ structure functions and
find that the calculated structure function agrees with the H1 experimental data~\cite{13HERA2010}.

On the other hand, we study the relation between the mass of the constituent quarks and their distributions in the pion. The valence quark distributions in the pion with different quark mass are obtained. It is found that the valence quark distributions become narrower when its mass is growing larger. The mechanism behind this behavior needs to further studies.

Finally, we would like to stress that, with the constituent quark model and thanks to the
 DGLAP equations, the obtained results on the pion quark distributions at $Q^2=20$ ${\rm GeV}^2$ and $F_2^{\pi}$ structure functions are fairly well agreement with the experimental measurements. Besides, the proposed mechanism here can be also expanded to study other hadrons such as kaon and proton. Considering abundant experimental data of the proton PDFs, it is expected that one can able to have a better check of the method with calculating the proton PDFs. This work
constitutes a first step in this direction.

\begin{acknowledgments}

We would like to thank professor Jialun Ping for valuable discussions and suggestions.
This work is supported by the Strategic Priority Research Program of Chinese Academy of Sciences under the Grant NO. XDB34030301, the National Natural Science Foundation of China under the Grant NOs. 12005266, 12075288, 11735003, 11961141012 and Guangdong Major Project of
Basic and Applied Basic Research No. 2020B0301030008. It is also
supported by
the Youth Innovation Promotion Association CAS.

\end{acknowledgments}

\end{document}